\documentstyle[12pt]{article}

\input{epsf}

\font\cmr=cmr7

\newcommand{\be}{\begin{equation}}
\newcommand{\ee}{\end{equation}}
\newcommand{\bea}{\begin{eqnarray}}
\newcommand{\ena}{\end{eqnarray}}
\newcommand{\sect}[1]{\setcounter{equation}{0}\section{#1}}

\newcommand{\ga}{{\gamma}}
\newcommand{\om}{{\omega}}
\newcommand{\D}{{\Delta}}
\newcommand{\slP}{\raise.15ex\hbox{$/$}\kern-.53em\hbox{$P$}}
\newcommand{\slR}{\raise.15ex\hbox{$/$}\kern-.53em\hbox{$R$}}
\newcommand{\slL}{\raise.15ex\hbox{$/$}\kern-.53em\hbox{$L$}}

\begin{document}
\renewcommand{\thefootnote}{\fnsymbol{footnote}}
\newpage
\pagestyle{empty}
\setcounter{page}{0}

%%%%%%%%%%%%%%%%%%%%%%%%%%%%%%%%%%%%%%%%%%%%%%%%%%%%%%%%%%%%%%%
%%%%%%%%%%%%%%%%%%%% LOGO ENSLAPP - DEBUT  %%%%%%%%%%%%%%%%%%%%
%%%%%%%%%%%%%%%%%%%%%%%%%%%%%%%%%%%%%%%%%%%%%%%%%%%%%%%%%%%%%%%
\def\logoenslapp{\logolight}
%
%% NB!!! POUR GAGNER DU TEMPS OU ENVOYER A L'EXTERIEUR,
%% ZAPPER LE LAPIN
%% COMMENTER LA LIGNE DE COMMANDE SUIVANTE POUR ZAPPER LE LAPIN
%
%\def\logoenslapp{\input epsf \logolapin}
%
%%%%%%%%%%%%%%%%%%%%%%%%%%%%%%%%%%%%%%%%%%%%%%%%%%%%%%%%%%%%%%%
%
\newcommand{\norm}[1]{{\protect\normalsize{#1}}}
\newcommand{\LAP}
{{\small E}\norm{N}{\large S}{\Large L}{\large A}\norm{P}{\small P}}
\newcommand{\sLAP}{{\scriptsize E}{\footnotesize{N}}{\small S}{\norm L}$
${\small A}{\footnotesize{P}}{\scriptsize P}}
\def\logolight{{\bf{{\large E}{\Large N}{\LARGE S}{\huge L}{\LARGE
        A}{\Large P}{\large P}} }}
\begin{minipage}{5.2cm}
  \begin{center}
    {\bf Groupe d'Annecy\\ \ \\
      Laboratoire d'Annecy-le-Vieux de Physique des Particules}
  \end{center}
\end{minipage}
\hfill
\logoenslapp
\hfill
\begin{minipage}{4.2cm}
  \begin{center}
    {\bf Groupe de Lyon\\ \ \\
      Ecole Normale Sup\'erieure de Lyon}
  \end{center}
\end{minipage}
\\[.3cm]
\centerline{\rule{12cm}{.42mm}}
%%%%%%%%%%%%%%%%%%%%%%%%%%%%%%%%%%%%%%%%%%%%%%%%%%%%%%%%%%%%%%%
%%%%%%%%%%%%%%%%%%%%% LOGO ENSLAPP  - FIN %%%%%%%%%%%%%%%%%%%%%
%%%%%%%%%%%%%%%%%%%%%%%%%%%%%%%%%%%%%%%%%%%%%%%%%%%%%%%%%%%%%%%

\vspace{20mm}

\begin{center}

{\LARGE {\bf  Some aspects of thermal field theories}\footnote{Talk
presented at the Fourth Workshop in High Energy Physics Phenomenology
(WHEPP 4), Calcutta, January 1996}}\\[1cm]

\vspace{10mm}

{\large P.~Aurenche$^{1}$}\\[.42cm]

{\em Laboratoire de Physique Th\'eorique }\LAP\footnote{URA 14-36 
du CNRS, associ\'ee \`a l'Ecole Normale Sup\'erieure de Lyon et
\`a l'Universit\'e de Savoie.}\\[.242cm]

$^{1}$ Groupe d'Annecy: LAPP, BP 110, F-74941
Annecy-le-Vieux Cedex, France.

%$^{2}$ Groupe de Lyon: ENS Lyon, 46 all\'ee d'Italie, F-69364 Lyon
%Cedex 07,France.
%\\

\end{center}
\vspace{20mm}

\centerline{ {\bf Abstract}}

\indent

Different formalisms used in the perturbative approach to Thermal Field
Theory (TFT) are briefly reviewed. The rate of production of a virtual photon in
a quark-gluon plasma is then discussed to illustrate some
features of TFT before introducing the Hard Thermal Loop resummation
which improves the behaviour of the theory in the infrared sector. Some
of the successes of this appraoch are described before turning to the
problems associated to the undamped transverse gluon oscillations and
the light-cone singularities arising when on-shell and/or massless particles
are involved.

\vfill
\rightline{hep-ph/9612432}
\rightline{\LAP-A-624/96}
\rightline{November 1996}

\newpage
\pagestyle{plain}
\renewcommand{\thefootnote}{\arabic{footnote}}

\sect{Introduction}

Thermal Field Theories (TFT) represent a very wide subject and only a few
aspects will be touched upon in these notes. The physical systems
considered will be the hot QCD (quark-gluon) plasma or the hot
QED (electron-photon) plasma in equilibrium. To probe the
properties of the plasma one often studies rare processes in the 
plasma which do not disturb the thermal equilibrium. As an illustration
of such processes we shall consider the production of real or virtual 
(lepton pair) photons in a QCD plasma. The rate of production is
proportionnal to $e^2$, where $e$, the electric charge, is taken to be
$e \ll g$, with $g$ the strong interaction coupling which is characteristic of
forces maintaining the plasma in equilibrium.

The following discussion will be based on perturbation theory and its
improvements. The main problem one encounters in such an approach
is that of divergences, both of the infra-red and the collinear types,
related to the masslessness of quarks and gluons.  
First the Feynman rules at finite temperature will be introduced and
several versions of them will be presented depending on the formalism
used: imaginary-time formalism (ITF), real-time formalism (RTF),
retarded/advanced (R/A) formalism. The production of a virtual photon in
a QCD plasma will be discussed and contrasted with the case at zero
temperature. The improvement of perturbation theory when soft momenta
are involved is presented: this is the hard thermal  loop (HTL)
resummation of Braaten-Pisarski and Frenkel and Taylor and some
consequences are discussed. Further problems associated to  mass
singularities are discussed and the need to go beyond the HTL scheme
is stressed.

\section{Feynman rules}

Thermal field theory is the application of the technics of usual quantum
field  theory which describes the interactions among a few fundamental
``particles", to the study of systems characterized by a large number of
particles (statistical systems) at a given temperature (thermal
equilibrium is assumed) \cite{FetteW1,Kapus1,Mills1}. The thermal expectation value of an operator
$A$ is defined by:
\bea
<A>_\beta &=& Z^{-1} \hbox{\rm Tr}( e^{-\beta H} A) \nonumber \\
&=& Z^{-1} \sum_{m} e^{-\beta E_m} <m|A|m>
\ena
where the partition function is $ Z = \hbox{\rm Tr}(e^{-\beta H}) $,
with $\beta= 1/T$ the inverse of the temperature. If one remembers the
time evolution for an operator in the Heisenberg representation
\be
e^{i H t'} A(t)\ e^{- i H t'} = A(t+t')
\ee
then one can write 
\be
e^{\beta H } A(t)\ e^{- \beta H } = A(t- i \beta)
\ee
and interpret the inverse temperature as an imaginary time. In TFT
it is then natural to introduce a complex time variable,
the imaginary part of which being related to the temperature. Consider
now the thermal expectation value of a bilocal operator
\bea
<A(t) B(t')>_\beta &=& Z^{-1} \hbox{\rm Tr}\left( e^{-\beta H} 
A(t) B(t') \right) \nonumber \\
&=& Z^{-1} \hbox{\rm Tr}\left( e^{-\beta H} e^{\beta H} B(t') 
e^{-\beta H} A(t) \right) \nonumber \\
 &=& < B(t'-i\beta) A(t)>_\beta
\label{eq:kms}
\ena
where the property of the cyclicity of the trace has been used.
This equation summarizes the important Kubo-Martin-Schwinger
(KMS) condition which expresses thermal equilibrium. Projecting 
on complete sets of states and introducing the evolution operators, the 
correlation function can be written as
\bea 
<A(t) B(t')>_\beta &=& Z^{-1}\sum_{m} e^{-\beta E_m} <m|A(t) B(t')|m>
\nonumber \\
&=& Z^{-1}\sum_{m,n} e^{-i E_n(t-t')}e^{i E_m (t-t'+i\beta)}
<m|A(0)|n><n| B(0)|m>.
\ena
For this expression to be defined, the exponentials should be well
behaved when $ E_m \rightarrow \infty$ and this requires
\be 
- \beta \le \mbox{\rm Im}(t-t') \le 0.
\label{eq:contour}
\ee
We turn now to a scalar field theory where
$\phi(x)$ is a real field. The propagator is
\bea
<G_c(x-x')>_{\beta} &=& <\theta_c(t-t') \phi(x) \phi(x') +
\theta_c(t'-t) \phi(x') \phi(x) >_{\beta} \nonumber \\ 
&=& < \theta_c(t-t')\ G^+(x-x') + \theta_c(t'-t)\ G^-(x-x') >_{\beta}.
\label{eq:propa}
\ena
Remembering that the time variable can be complex, the time ordering
operator $T_c$, implied by the function $\theta_c(t-t')$, generalizes
on an oriented contour $C$ in the complex time plane, the time ordering
operator defined on the real axis.
%and this funtion is unity if $t$ precedes $t'$ on the contour and zero
%otherwise. 
Then, according to eq.~(\ref{eq:contour}) the oriented contour has to be
descending. The application of the KMS condition to the thermal
propagator leads to the relations
\bea 
G^-(t-t') = G^+(t - i\beta -t')\ \ {\mbox {\rm and} }\ \
G^+(t-t') = G^-(t -t' + i\beta). 
\ena
One obtains the propagator and other Green functions from the 
generating functional
\bea
Z[j] = \mbox{\rm Tr} \left( e^{-\beta H} T_c 
\exp \left( i \int_c d^4 x j(x) {\hat \phi}(x) \right) \right)
\ena
by the usual differentiation formula
\be 
G_c(x_1, \cdots, x_n) = {1 \over Z[0]} 
{\delta^n \ln(Z[j]) \over i\delta j(x_1) \cdots i\delta j(x_n)}
\ee
Introducing the eigenvalue of the field operator $\hat \phi$
at some initial time $t_i$
\be
{\hat \phi}(x) |\phi (\vec x,t_i)> = \phi(\vec x) |\phi (\vec x,t_i)>
\ee
one has to evaluate
\bea
Z[j] &=& \int d \phi <\phi,t_i| e^{-\beta H} T_c 
\exp \left( i \int_c d^4 x j(x) \phi (\vec x) \right) |\phi,t_i> 
\nonumber  \\
&=& \int d \phi <\phi,t_i - i\beta|  T_c 
\exp \left( i \int_c d^4 x j(x) \phi (\vec x) \right) |\phi,t_i>. 
\label{eq:genfun} 
\ena
Using well-known technics of splitting the time interval in small 
intervals one re-writes the generating functional as a functional
integral 
\be
Z[j] =  N \int [{\cal D}(\phi)] \exp i \int_c d^4 x 
\left( {\cal L}(x) + j(x) \phi(x) \right)
\ee
where the time contour runs from $t_i$ to $t_i - i \beta$ and the
fields are subject to the periodic boundary condition
\be
\phi(t_i,\vec x) = \phi(t_i - i \beta,\vec x) 
\ee
because of the same ``initial" and ``final" state as implied by the
Trace operation. From now on, one constructs the perturbative expansion,
as in the zero temperature case, by decomposing the lagrangian
into its free part and its interacting part.

We concentrate now on the construction of the propagator before
specifying the contour which will lead to the different formalisms
mentioned in the introduction. First, introduce the Fourier transform
of the functions $G_c^\pm$
\be
D^\pm(K) = \int d^4 x e^{iKx} G_c^\pm(x).
\ee
The KMS conditions easily give the relation
\be
D^-(K)=e^{\beta k_0} D^+(K)
\ee
and defining now the spectral function $\rho(K)=D^+(K)-D^-(K)$
one derives
\bea 
D^-(K) &=& \rho(K) n_{_B}(k_0), \qquad \mbox{\rm with } n_{_B}(k_0)=
{1\over e^{\beta k_0} -1}   \nonumber \\
D^+(K) &=& \rho(K) ( n_{_B}(k_0) +1)
\ena
and in the space-time representation
\bea
G_c(x) = \int {d^4 K \over (2 \pi )^4} e^{-iKx} 
\rho(K) (\theta_c(t)+n_{_B}(k_0)).
\ena
To obtain the free field propagator one expands the field into the
creation and annihilation operators: 
\be
\phi(x) = \int {d^4 K \over (2 \pi )^4} \ 2 \pi \theta(k_0) 
\delta(K^2 -m^2) \left[ a(K) e^{-iKx} + a^{\dagger}(K) e^{iKx} \right]
\ee
and one constructs the propagator from its definition eq. 
(\ref{eq:propa}). Making use of the usual commutation relations, 
$e.\ g.$,
\be
[a(K),a^{\dagger}(K')] = (2 \pi)^3\ 2 \omega_k\ \delta(\vec k -\vec k')  
\ee
one finds the general expression
\be
G_c(x-y) = \int {d^4 K \over (2 \pi )^3} e^{-iK(x-y)} 
\epsilon(k^0) \delta(K^2 -m^2) (\theta_c(x^0 - y^0)+n_{_B}(k_0)).
\label{eq:propa2}
\ee
One immediately obtains the spectral function of the free propagator
to be
\be
\rho(K) = 2 \pi \epsilon(k^0) \delta(K^2 -m^2).
\ee
To proceed further with actual calculations it is necessary to choose a
contour and this will lead to the different formalisms with different
expressions for the Green's functions. Of course, the predictions for
rates or cross sections should be independent of the choice of the
contour.

\subsection{The imaginary-time formalism}
This is based on the simplest contour, running vertically in the time
interval $[0,-i\beta]$ \cite{FetteW1,Kapus1,LandsW1}. It is convenient
to introduce the ``imaginary" time $\tau$, related to the ``real" time
by $t = - i \tau$. Using the KMS conditions one shows that the
propagator is a periodic function of the energy of period 
$2 \pi i /\beta$. The propagator is expanded as a Fourier series with
the Matsubara frequencies $\omega_n = 2 \pi n / \beta$ and one has
\bea
G(\tau,\vec x) = {i \over \beta} \sum^\infty_{n=-\infty}
\int {d^3 k \over (2 \pi )^3} e^{-i\omega_n \tau} 
 e^{i\vec {k}\vec {x}} \Delta (i\omega_n, \vec x) 
\ena
or equivalently
\bea
\Delta (i\omega_n, \vec x) = \int^\beta_0 d \tau \int d^3 x 
e^{i\omega_n \tau}  e^{-i\vec {k}\vec {x}} G(\tau,\vec x). 
\ena
Using eq. (\ref{eq:propa2}) with $\theta_c(x^0 - y^0)=1$ when
$0 \le \tau \le \beta$ one finds:
\begin{itemize}
\item[(i)] propagator: 
$$\Delta (i\omega_n, \vec k) =
{i \over (i\omega_n)^2 - \omega_k^2} \qquad \qquad \mbox{\rm with} 
\qquad \omega_k^2={\vec k}^2+m^2;$$
\item[(ii)] vertex: $\qquad\qquad\qquad  -ig$ as at $T=0$;
\item[(iii)] energy-momentum conservation: $(2 \pi)^4 \delta(K) 
\rightarrow -i \beta \delta_{n,0} (2 \pi)^3 \delta(\vec k);$
\item[(iv)] loop integral: 
$$\int {d^4 K \over (2 \pi )^4} \rightarrow {i \over \beta}
\sum^{n=\infty}_{n=\infty} 
\int {d^3 k \over (2 \pi )^3}\qquad\qquad\qquad.$$
\end{itemize}
When working in the ITF an analytical continuation to real external
energy variables must be performed 
\be
i\omega_n \rightarrow k^0 \pm i\varepsilon, \qquad 
\varepsilon \rightarrow 0_+ ,
\label{eq:retadv}
\ee
corresponding to ``retarded" and ``advanced" energies.

\subsection{The real-time formalism}

The idea is to include the real axis in the oriented time contour as
shown in Fig.~1 \cite{Mills1,LandsW1}. 
%The contour runs from some
%initial time $t_i$ to some final time $t_f$ then goes down vertically to
%$t_f- i \sigma$ and horizontally back to $t_i- i \sigma$ and finally
%vertically to  $t_i- i \beta$. 
The arbitrary parameter $\sigma$ is such
that $0\le \sigma \le \beta$. It is convenient to introduce the following
propagators depending of the relative position of $x$ and $y$ (see 
Fig.~1 for the label of the various pieces of the contour):

\begin{itemize}
\item[(i)]
if $x,\ y$ on $C_1\qquad \qquad  G(x-y) = G^{11}(x-y)$
\item[(ii)]
if $x,\ y$ on $C_2\qquad \qquad  G(x-y) = G^{22}(x-y)$
\item[(iii)]
if $x$ on $C_1$, $y$ on $C_2\ \ \ \ G(x-y) = G^{12}(x-y)
=G^{-}(x^0-y^0+i\sigma, \vec x-\vec y)$
\item[(iv)]
if $x$ on $C_2$, $y$ on $C_1\ \ \ \ G(x-y) = G^{21}(x-y)
=G^{+}(x^0-i\sigma-y^0, \vec y-\vec x)$
\end{itemize}
The cases when one (or both) of the time coordinates is on a vertical
part of the contour are not explicitely needed when working in 
momentum space although the vertical pieces of the contour play a crucial 
role in deriving a consistent formalism \cite{Evans6,Gelis1}. From the general expressions
above one obtains the Fourier transforms:
\bea
D^{11}(K)&=&\ {i \over K^2 - m^2 + i\varepsilon} + 2 \pi n_{_B}(|k^0|)
\ \delta(K^2 - m^2) \nonumber \\ 
D^{22}(K)&=& - {i \over K^2 - m^2 - i\varepsilon} + 2 \pi n_{_B}(|k^0|)
\ \delta(K^2 - m^2) = (D^{11}(k))^* \nonumber \\ 
D^{12}(K)&=&\  2 \pi (\theta(-k^0)+n_{_B}(|k^0|)\ \delta(K^2 - m^2) 
\ e^{\sigma k^0} \nonumber \\ 
D^{21}(K)&=&\  2 \pi (\theta(k^0)+n_{_B}(|k^0|)\ \delta(K^2 - m^2) 
\ e^{-\sigma k^0}. 
\label{eq:rules}
\ena
Similar formulae can be derived for a fermion and they involve the
Fermi-Dirac fonction $n_{_F}(|k^0|) = 1/ (e^{\beta |k_0|} + 1)$ 
rather than $n_{_B}(|k^0|)$.
A nice feature of the RTF is that it separates the temperature 
dependence contained in the statistical factors from the $T=0$ part. 
In the RTF there is effectively a ``doubling" of fields since one 
 may introduce 
\bea
\phi^1(x) &\equiv& \phi(x)\ \ \mbox{when }x\mbox{ is on the horizontal
contour } C_1 \nonumber \\
\phi^2(x) &\equiv& \phi(x-i \sigma)\ \ \mbox{when }x\mbox{ is on the horizontal
contour } C_2. 
\ena
This leads to two types of vertices: fields of type $\phi^1$
couple together with strength $-ig$ or fields of type $\phi^2$
couple with strength $ig$, the change of sign in the latter case
arising because the time
variable is running from right to left. The propagator in the
RTF is a $2 \times 2$ matrix ${\cal D}(k) =  \left(D^{ij}(k)\right)$
with elements defined in eqs. (\ref{eq:rules}). When calculating
Green's functions in this formalism it is necessary to sum over ``type 1"
and ``type 2" internal vertices. This is crucial to cancel mathematically
undefined terms, such as products of $\delta$-functions with the same 
argument, which appear in intermediate stages of the calculation: to 
illustrate this point it is sufficient to consider self-energy insertions
on a propagator and check that, only after summation over all types of
internal vertices, terms of the form $(\delta(K^2-m^2))^n$ disappear.

``Cutting rules" for calculating the imaginary part of Green's functions
have been derived which are extremely useful to obtain rates or cross
sections \cite{KobesS1,KobesS2}.  The relation between RTF Green's
functions and the analytic continuation of ITF ones is neither trivial
nor simple \cite{Kobes2,Kobes1,Evans5,Gueri2} and has led to some
confusion in the past \cite{BaierPS5,NakkaNY3}. A nice way to relate the
two, in some cases, is described next.

\subsection{The R/A formalism}

This formalism is a real-time formalism but it allows to construct
directly retarded/advanced Green's functions, $i.e.$ functions with
external  energies ($k_0+i\varepsilon$) or  ($k_0-i\varepsilon$) which
are related to the ITF Green's functions analytically continued as in
eq.~(\ref{eq:retadv}) \cite{AurenB1,EijckW1}. The basic step is the
following decomposition of the RTF propagator
\bea
{\cal D}(K) = U(K)
\left( \begin{array}{cc}
\Delta_{R}(K) & 0 \\
0 & \Delta_{A}(K) \end{array} \right) V(K)
\label{eq:RA}
\ena
where the retarded and advanced propagators are given by
\be
\Delta_{_{R,A}}(K)\equiv{{i}\over{K^2-m^2\pm i \varepsilon k^0}},
\ \ \ \varepsilon \rightarrow 0.
\ee
The matrices $U$ and $V$ depend on momentum $K$ as well as the
thermal factor and the parameter $\sigma$. They are associated to the
RTF vertices and after summation on the two types of vertices one is led
to the following momentum  dependent vertices 
\bea
&&g_{_{AAA}}(P,Q,R)=g_{_{RRR}}(P,Q,R)=0, \nonumber\\
&&g_{_{RRA}} (P,Q,R) = g_{_{ARR}} (P,Q,R) =g_{_{RAR}} (P,Q,R) =g,
  \nonumber\\
&&g_{_{RAA}}(P,Q,R) = - g\ (1+n_{_{B}}(q^0)+n_{_{B}}(r^0)), \nonumber\\
&&g_{_{ARA}} (P,Q,R) = -g\ (1+n_{_{B}}(p^0)+n_{_{B}}(r^0)),\ \cdots 
\end{eqnarray}
where the indices refer to the retarded/advanced prescriptions for the
incoming external momenta $P,\ Q$ and $R$. The first two equalities
above express the KMS conditions in the R/A formalism.
In loop calculations the sum over $R$ and $A$ internal indices is
implied. Some useful discontinuity formulae to calculate cross sections
have been derived in this formalism \cite{AurenB1,AurenBP1}.

\section{Lepton pair production in a QCD plasma in perturbation theory}

It is instructive to consider first the production of a lepton pair in a
QCD plasma: this illustrates some of the features and problems occuring
in thermal calculations \cite{CleymD2,BaierPS2,AltheAB1}.  When
calculating this rate,  at the first order in QCD, in hadronic
collisions ($i.e.$ at $T=0$) it is well known that,  at the partonic
level, the infrared divergences (soft gluon) cancel according to the
Lee-Kinoshita-Nauenberg theorem while the collinear  divergences survive
due to the collinear emission of a gluon by the  annihilating quarks
(process $q \bar q \rightarrow g \gamma^*$) or the collinear splitting
of an initial gluon in the process $g q \rightarrow q \gamma^*$.
However, the partons are confined and these divergences are absorbed,
via the factorization theorem, in a re-definition of the structure
functions of the hadrons which become scale dependent: in other words
confinement shields  these singularities. Since, in a quark-gluon
plasma, the partons are deconfined, it is legitimate to ask what happens
to the collinear  singularities.

For a pair of invariant mass squared $Q^2$, produced at rest, the rate 
is \cite{GaleK1}: 
\be
\frac{d N}{d q_0 d^3 q d^4x} = - \frac{\alpha }{ 12 \pi^3} 
\frac{1}{Q^2} n_{_{B}} (q_0)
  \ \hbox{\rm Im}\, \Pi^\mu\,_\mu (Q).
\label{ratevir}
\end{equation}
where $\hbox{\rm Im}\, \Pi^\mu\,_\mu (Q)$ is the imaginary part of
the trace of the retarded photon self energy. It has been calculated at
the first order in QCD in the RTF and the result was found to be 
finite, of the form \cite{AltheA1,GabelGP1}:
\be
\hbox{\rm Im}\, \Pi^\mu\,_\mu (Q) = {\alpha \over \pi} N_{_C}
\left(1 - 2 n_{_F}({Q\over2}) \right) Q^2 \left[ 1 + 
{\alpha_s \over \pi} c_{_F} \left({3 \over 4} + F({Q\over T}) \right)
\right].
\label{eq:corr}
\ee
In the above equation we have considered only one species of massless
quarks of charge $e$; $\alpha_s$ is the fine structure constant of the
strong interactions and $N_{_C}$ and $c_{_F}$ are the usual colour
factors. Except for an overall normalisation factor, the thermal
dependence is contained in the function $F$. The  calculation shows that
both infrared and collinear divergences cancel in the $T=0$ as well as
in the thermal piece \cite{CleymD2,BaierPS2,AltheAB1}. 
This occurs for two reasons: the phase space
available at finite temperature is larger than at 0 temperature 
(one has to integrate over initial particle momenta) and
processes such as $q \bar q g \rightarrow  \gamma^*$ are included
together with the usual annihilation and Compton diagrams of the
Drell-Yan process (in this respect the factor $3/4$ in the correction
term of eq.~(\ref{eq:corr}) is exactly the factor appearing in the
$T=0$ annihilation of $e^+ e^- \rightarrow q \bar q g$); secondly there exist
relations among the Bose and Fermi statistical factors which express
thermal equilibrium. For example, in equilibrium the rate of transition
for a gluon of energy $E_1$ to be absorbed by a fermion of energy $E_2$
is the same as the rate for a fermion of energy $E_1+E_2$ to decay into a
gluon of energy $E_1$ and a fermion (detailed balance principle). 
This leads to relations of type
\be
n_{_B}({E_1})n_{_F}({E_2}) = n_{_F}({E_1+E_2})
(1 + n_{_B}({E_1}) - n_{_F}({E_2}))
\ee
and other similar relations. The cancellation of divergences has been
explicitely shown to hold in scalar theories up to 3 loops
\cite{GrandBP2} but no general proof exists (see however \cite{Niega3}).
Recently it has been shown that the cancellation of singularities can
also occur out of equilibrium \cite{BellaM2}. The thermal correction
factor has been calculated and it behaves as  \cite{AltheA1}
\bea
{\alpha_s \over \pi} F({Q\over T}) &\equiv& g^2 {T^2 \over Q^2}, 
\qquad \mbox{\rm as } {T^2 \over Q^2} \rightarrow 0  \nonumber \\
{\alpha_s \over \pi} F({Q\over T}) &\equiv& g^2 {T^2 \over Q^2} 
\ln{T^2 \over Q^2}, \qquad \mbox{\rm as } {T^2 \over Q^2} 
\rightarrow \infty.
\label{eq:finite}
\ena
This result can be understood as follows: the $T^2$ factor reflects
the size of the thermal phase space while the $1/Q^2$ is necessary for
dimensional reasons. When the lepton pair invariant mass becomes of order
$g T$ the perturbation series appears not to be well behaved
as the correction term becomes as large as the lowest order one
and when $Q^2 \rightarrow 0$ a divergence appears again.
This illustrates a general feature of thermal field theories where
perturbation theory breaks down when ``soft" scales, $i.e.$ scales
of order $g T$, are involved. This is put in a rigorous way in the
``Hard Thermal Loop" (HTL) resummation scheme of Braaten and Pisarski
\cite{BraatP1,BraatP2}  and Frenkel and Taylor \cite{FrenkT1,FrenkT2} to
which we turn next. In this approach, it is  natural to distinguish
between ``hard" scales, of ${\cal O}(T)$, the typical energies of quarks
and gluons in the QCD plasma, from the soft scales, of ${\cal O}(gT)$,
with the assumption $g \ll 1$.

\section{Hard Thermal Loop Resummation}

Consider, as an example, the retarded self-energy of a quark 
of momentum $P$ in a QCD plasma. At the one-loop order, it can be 
expressed as 
\bea
-i\Sigma_{R} (P) &=& c_{_F} g^2 \int {d^4L\over(2\pi)^3} \ \ga_\nu 
\ (\slP+\slL)\ 
\ga^\nu  \left[({1\over2}+n_{_B}(l_0))\ \epsilon (l^0) \delta (L^2)
\ \D_{R} (P+L) \nonumber \right.\\
& &\ \ \ \ \left. + ({1\over2}-n_{_F}(p_0+l_0))\ \epsilon (p^0+l^0) 
\delta ((P+L)^2)\ \D_{A}(L)\right]. 
\label{eq:self}
\ena
If the external momentum $P$ is soft, with components of ${\cal O}(g T)$,
this expression reduces to
\be
\Sigma_R(P) \sim { c_{_F} g^2\over 4\pi^2} \int d \om\ \om
\ (n_{_B}(\om) +n_{_F}(\om)) \int {d\hat L\over 2\pi} 
\ {\hat{\slL}\over P\hat L+i\varepsilon} 
\label{eq:self2}
\ee
where we have introduced the light-like vector $\hat L = (1,\hat l)$
and $\omega = |l_0|$.
The  dimensional and the angular part of the loop integration factorize
and it comes out \cite{BraatP1}:
\be
\Sigma_R(P) \sim {m^2_{\hbox{\cmr q}} \over 2} \int {d\hat L\over 2\pi}
\ {\hat{\slL} \over P\hat L +i\varepsilon}.
\label{eq:self3}
\ee
The factor $m^2_{\hbox{\cmr q}}$ results from the $\omega$ integration 
and it is 
\be
m^2_{\hbox{\cmr q}} = c_{_F} {g^2 T^2\over 8}. 
\label{eq:mass}
\ee
The $T^2$ behavior arises entirely from the region where $\om \sim T$,
$i.e.$ when the loop momentum is ``hard", the ``soft" contribution of
the loop being suppressed by  factors of $g$. When
inserting the self-energy correction on the fermion propagator the effective
propagator is obtained:
\be
^*S_R (P) = {i\over\slP-\Sigma_R(P)}
\label{eq:effpro} 
\ee
It is clear from eq. (\ref{eq:self3}) that when $P$ is soft the
self-energy correction is also ${\cal O}(gT)$ and the modification
implied by the loop is of the same order as the bare propagator. On the
contrary when $P\sim T$ the self energy insertion induces a correction
of ${\cal O} (g)$. As a consequence, for a consistent perturbative
calculation, one should use the effective propagator
eq.~(\ref{eq:effpro}) when the momentum is soft while the bare one is
sufficient when the momentum is hard. The pole in the propagator
eq.~(\ref{eq:effpro}) leads to complicated dispersion relations  with
two branches (quasi-particle excitations) and the fermion acquires an
effective mass $m_{\hbox{\cmr q}}$ at rest given by eq.~(\ref{eq:mass});
for hard momentum, one of the dispersion relations \cite{Klimo1,Weldo2}
leads to a 
particle-like excitation of mass $2 m_{\hbox{\cmr q}}$ whereas the other
branch (called plasmino) decouples exponentially. An important
consequence of thermal corrections is that the self energy acquires an
imaginary part in the space-like region as can be seen by evaluating
explicitely
\bea
\mbox {\rm Im} \Sigma_{R} (P) &=& -  {m^2_{\hbox{\cmr q}} \over 4}  
\int d \hat L \hat{\slL} \delta (P \hat L)
\label{eq:imsig}
\ena
If $P^2 < 0$ then one immediatly finds $\mbox {\rm Im} \Sigma \sim g T$
while if  $P^2 > 0$ it is vanishing, as the $\delta$-function has no
support (an exact calculation would give $g^2 T$), and the
quasi-particle interpretation is justified. The origin of the imaginary
part when $P^2 < 0$ is the  Landau damping effect which accounts for the
scattering of quarks and gluons in the thermal bath. 
%which is due to the absorption or emission 
%of a hard fermion in the plasma by hard gluon to become a soft fermion. 
Such a mechanism has no equivalent at zero temperature where the
imaginary part exists only above the threshold $P^2 > 0.$

In the same way one finds that the gluon propagator has a hard thermal loop
contribution. In fact, in the Landau gauge, the effective gluon 
propagator takes the form \cite{Klimo1,Weldo1} (for simplicity, we do not
specify the R/A indices)
\bea
^* D^{\mu\nu}(L)\equiv - {P^{\mu\nu}_{_{T}}(L)}\Delta^{^{T}}(L)
-{P^{\mu\nu}_{_{L}}(L)}\Delta^{^{L}}(L)
\label{eq:effglu}
\ena
where
\bea
\Delta^{^{T,L}}(L)\equiv{i\over{L^2-\Pi_{_{T,L}}(L)}}
\label{eq:effglu2}
\ena 
The tensors $P^{\mu\nu}_{_{T,L}}(L)$ are the transverse and longitudinal
projectors whose explicit  expressions are found in
\cite{Weldo1}. The vanishing of the denominators implies
different dispersion relations for the transverse and longitudinal
polarisation states. One has the following limits:
\bea
\Pi_{_{T}}(\om,{\vec l}=0) = \Pi_{_{L}}(\om,{\vec l}=0)  
= m^2_{\hbox{\cmr g}} =  g^2 T^2 [N+N_{\hbox{\cmr f}}/2]/9     
\label{eq:massgl}
\ena
which means that the gluon acquires an effective mass in the plasma.
On the other hand, in the static limit,
\bea
\Pi_{_{T}}( \om =0, {\vec l} \rightarrow 0) &=& 0 
\label{eq:dampt} \\
\Pi_{_{L}}( \om =0, {\vec l}  \rightarrow 0)   &=& 3  m^2_{\hbox{\cmr g}} 
\label{eq:dampl}
\ena
implying static screening of the electric field  but not of the 
magnetic field.

Other Green's functions may receive hard loop corrections. For example,
for the quark-quark-photon vertex (which we will need later) the HTL
contribution is simply written as \cite{BraatP1}
\bea
- i e V_\lambda{}(P,Q,R) = - i e {m^2_{\hbox{\cmr q}}\over2}  \int 
{d\hat L\over 2\pi} \ { \hat L_\lambda \hat{\slL}
\over P\hat L \ R\hat L }  
\label{eq:vertex}
\ena
where $Q$ is the photon momentum and $\hat L = (1,\hat l)$ is as in
eq.~(\ref{eq:self2}). When the external momenta $P, Q$ and $R$ are soft
the  integrand is $\sim 1/(g T)^2$ and the one loop vertex is  of
${\cal O}(e)$ as the bare QED vertex. On the contrary, if at least one
of the external lines is hard the effective vertex is suppressed by at
least one power of $g$. One is naturally led to introduce an effective 
vertex
\be
\tilde \Gamma_\lambda (P,Q,R) =  
- i e (\gamma_\lambda + V_\lambda{}(P,Q,R) )
\label{eq:effver}
\ee
to be used, instead of the bare vertex, whenever all $P,\ Q$ and $R$
are soft.

Likewise it can be shown that the three-gluon vertex, and more generally
the $n$-gluon vertex receive hard thermal loop corrections, as do
Green's functions with 2-fermions and $n-2$ gauge bosons. Furthermore,
in the HTL approximation the hard thermal loops are gauge invariant and
they satisfy QED-like Ward identities \cite{BraatP1}. For example, for
the 3-gluon and the $q-\bar q-g$ vertex, respectively,
\bea
{R^\lambda}\ ^*\Gamma_{\lambda \mu \nu}(P,Q,R) &=& 
-\ ^*D^{-1}_{\mu\nu}(P) +\ ^*D^{-1}_{\mu\nu}(P) \nonumber \\
{Q^\lambda}\ ^*{\tilde \Gamma}_{\lambda}(P,Q,R) &=& 
- i ( ^*S^{-1}(P) + ^*S^{-1}(R) )
\label{eq:ward}
\ena						
In conclusion, when carrying out a perturbative calculation, to obtain a
consistent result bare soft lines and vertices should be replaced by
their effective counterparts: resummed propagators defined in
eqs.~(\ref{eq:effpro},\ref{eq:effglu}) and  effective vertices  ($e.g.$
eq.~(\ref{eq:effver})). In Feynman diagrams effective Green's functions
will be indicated by a $\bullet$ to distinguish them from their bare
counterparts. Two-loop corrections are not needed in this scheme since
internal lines in self-energy and, more generally $n-$point functions,
being hard, corrections to these will be suppressed by factors of $g$.
These rules can be deduced systematically from an effective 
gauge-invariant lagrangian \cite{TayloW1,Braaten}. 
The application of this scheme has lead to
tremendous progress in the calculation of observables and to the
solution of long standing puzzles. 
%such as that of the damping rate and also rendered the calculation of
%some processes finite due to an improved behaviour in the infrared
%sector. We briefly review some of the successes of the HTL approach
%before turning to open problems.

\section{Applications of the HTL resummation}

We start by considering again the production of a soft virtual photon at
rest \cite{BraatPY1}. As discussed before one should calculate the
imaginary part of the vacuum polarisation diagram which, at lowest
order, describes the annihilation of a soft quark-antiquark pair in a
plasma. Since all momenta in the problem are soft one should use
effective propagators and vertices and the graph to be  considered is
shown on Fig.~2. There exist other loop diagrams but, in
principle, they do not contribute when summing over the photon
polarisation indices.  In terms of scattering amplitudes the graph of
Fig.~2 has a very rich structure. Taking the imaginary part
implies cutting through the effective vertices and propagators.
Correspondingly, the rate of virtual photon production involves
convolution of the diagrams in Fig.~3 \cite{BraatPY1}:  if both internal
fermion lines are time-like ($P^2,\ R^2>0$) the photon is produced in
the annihilation or the decay of 2 quasi-particles (pole-pole) interaction
(a); if one of the fermion line is space-like the imaginary part of the
propagator corresponds to Landau damping and give rise to the scattering
process (pole-cut) in (b); if both fermion lines are space-like then we
obtain the double-scattering process (cut-cut) of (c). Cutting through
effective vertices lead to similar diagrams (interference of (c) and
(d)). The effect of Landau damping is of uttermost importance and
increases the rate of production by several orders of magnitude compared
to the estimates of eq.~(\ref{eq:corr}) \cite{BraatPY1}. \\ 
Similarly the rate of production of a hard photon has been calculated
and found to be finite in the HTL approach while it would be infinite at
lowest order of perturbation theory \cite{KapusLS1,BaierNNR1}: 
indeed, in the annihilation and
Compton processes of Fig.~4a, the fermion propagator may
diverge: for a static exchange, for example, one has to evaluate an
integral of the form $\int d p / p$ which diverges for soft 3-momentum
transfer $p$. In the HTL approach, the effective fermion propagator
eq.~(\ref{eq:self3})  should be used leading to the integral $\int d p \
8 p / (8 p^2 +m^2_{\hbox{\cmr q}}(4+\pi^2))$ regularized by the
fermion thermal mass. The rate is then 
\be q_0 {d R^\gamma \over d q^3} \sim  {\alpha \alpha_s \over 2 \pi^2} 
T^2 e^{-q_0 / T} \ln\left({q_0 T \over m^2_{\hbox{\cmr q}} } \right). 
\ee 
The calculation is much simpler than in the soft photon case as no
effective vertices is needed since the vertices are connected to two
hard lines: in fact it is sufficient to calculate the diagram of Fig.~4b
where one of the internal lines at least is hard since the photon
momentum is hard.  Many more applications of the resummation scheme to
the calculation of physical observables have been done \cite{Lebellac}.

There has been in the past much controversy concerning the gluon damping
rate ($i.e.$ the imaginary part of the gluon self-energy): a negative
rate would mean undamped oscillations and a perturbatively unstable
quark-gluon  plasma. Until the HTL scheme was developped  a wide range
of gauge dependent results could be found in the literature, based on
the one-loop approximation, with rates of different signs and strength
\cite{Pisarski}!
These results were in fact incomplete since they used only bare soft
propagators and vertices. Defining the damping rate of a transverse
gluon at rest as
$ \gamma_{_T}(0) = 
{\hbox {\rm Im}} \Pi_{_T}( m_{\hbox{\cmr g}},0) / 2 m_{\hbox{\cmr g}}$,
and performing a consistent calculation in the effective theory involves
the evaluation of the graphs of Fig.~5 with effective propagators and
vertices \cite{BraatP3}. The  result comes out gauge invariant 
and positive:
\be
\gamma_{_T}(0) = \gamma_{_L}(0) = 6.635\ {g^2 N_c \over 24 \pi} \ T
\ee
Among other successes one should mention the ${\cal O}(g^3)$ contribution
to eq.~(\ref{eq:massgl}) \cite{Schul1}.

\section{Problems in the HTL resummation}

The use of the effective theory of Braaten and Pisarski makes it
possible to improve the infrared behaviour of the theory in a gauge
invariant  manner. However there remains a number of difficulties
related to the lack of static screening of transverse gluon modes
(eq.~(\ref{eq:dampt}))  and/or the appearence of collinear singularities
when external particles are on-shell or massless. Two examples
will respectively illustrate this problem: the damping  rate of a fast
moving particle and the production of soft real photons.

Consider a fast moving fermion with energy $p_0 \gg T$. Its damping rate
is given in the effective approach by the imaginary part of the diagram
of Fig.~6 with the dominant contribution arising from a 
soft transverse gluon and a hard internal fermion. The result is a 
convolution of the transverse gluon spectral function $\rho_{_T}$
($i.e.$ the imaginary part of $\Delta^{^T}(L)$ in eq.~(\ref{eq:effglu2})
when $L^2 < 0$) and one finds
\bea
\gamma(P) \sim g^2 c_{_F} T \int_{\mbox {\cmr soft}} l d l
\int^l_{-l} {dl_0 \over l_0} \rho_{_T}(l_0,l) 
\sim g^2 c_{_F} T \int_{\mbox {\cmr soft}}  {d l \over l}.
\ena
This infrared divergence arises from unshielded static gluon exchanges
(see eq.~(\ref{eq:dampt})). It should be noted that this divergence
appears because one evaluates the diagram for an on-shell fermion, otherwise
the off-shellness of the external fermion would shield the infrared
singularity. In QCD this IR divergence can be
cured by the introduction of a magnetic mass (of non-perturbative origin)
of ${\cal O}(g^2 T)$ \cite{Rebha1,Pisar4}, a solution not applicable to QED where 
gauge invariance requires the magnetic mass to vanish. Several
proposals have been considered to solve the problem. An interesting 
recent solution is based on the observation that higher loop diagrams
with static transverse gluons contribute to the same order as the
graph of Fig.~6. Such corrections have been resummed
\cite{BlaizI2}, in
QED, by calculating the fermion propagator in a static background field
and, defining the damping rate as the inverse of the decay time of 
the propagator in space-time coordinates, one finds
\bea
\gamma \sim 
%g^2 T \ln ({m_{\hbox{\cmr q}} \over \gamma})  \sim 
g^2 T \ln ({1 \over g}).
\ena
Going back to momentum space the retarded fermion propagator 
$^*S_{_R}(p_0,{\vec p})$ appears to be an entire function with singularity
at ${\mbox {\rm Im}} p_0 \rightarrow - \infty$.

The origin of the mass singularity problem can already be
illustrated on a very simple example \cite{FlechR1}. Coming back to 
eq.~(\ref{eq:self3}) and constructing 
${\mbox {\rm Tr}}(\gamma^0 \Sigma_R(P))$ it comes out
\be
{\mbox {\rm Re\ Tr}}(\gamma^0 \Sigma_R(P)) \sim - {2 \over p}
m^2_{\hbox{\cmr q}} \ln \left( {p_0 -p \over p_0 +p} \right)
\label{eq:massing}
\ee
which diverges logarithmically for the massless on-shell condition. Such
a singular behaviour occurs also in the photon or gluon polarisation
tensor and higher order diagrams contribute to same order as the lowest
order one \cite{FlechR1}. Divergences like those in
eq.~(\ref{eq:massing})  appear when calculating the production
rate of soft real photons \cite{BaierPS1,AurenBP1}. The diagram to be
calculated is that of Fig.~2 and taking the imaginary part implies
``cutting" trough an effective vertex  (see an example in Fig.~7) $i.e.$
taking the imaginary part of eq.~(\ref{eq:vertex}) and therefore to
evaluate an integral of type:
\bea
{\mbox {\rm Im}} V_\lambda{}(P,Q,-R) \sim  {m^2_{\hbox{\cmr q}}\over2}  
\int  {d\hat L\over 2\pi} \ \hat L_\lambda \hat{\slL}
\ {\delta(P\hat L) \over R\hat L} 
\sim {m^2_{\hbox{\cmr q}}\over2}  
\int  {d\hat L\over 2\pi} \ \hat L_\lambda \hat{\slL}
\ {\delta(P\hat L) \over Q\hat L}   
\label{eq:softgam}
\ena
where the constraint $P=R-Q$ has been used in the last equation. For a
real photon, $Q$ is light-like as is $\hat L$ and therefore the above
integral exhibits a collinear singularity when $P$ and $R$ are collinear
since $P\hat L$ and $Q\hat L$ then vanish at the same point. Therefore,
when applying strictly the HTL approach and keeping bare propagators and
vertices when hard momenta are involved the production rate of soft real
photon in a QCD plasma appears not to be defined. Even more
interesting is the bremsstrahlung contribution to this rate
\cite{AurenGKP1,AurenGKP2} which is supposed to vanish in the HTL
approximation as it is contained in the gluon tadpole diagram: this
diagram summarizes in fact the graphs of Fig.~8,
where the internal momenta $P$ and $R$ are now hard and $L$ is soft,
and involves the evaluation of:
\be
\hbox{\rm Im}\,\Pi^\mu\,_\mu(Q) \sim e^2 g^2\ q_0
%\int \frac{d^4 P}{(2\pi)^3}  \int \frac{d^4 L}{(2\pi)^3}
\int d^4 P d^4 L \left({p \over l}\right)^2 n'_{_F}(p_0)
 n_{_{B}}(l_0)\rho_{_{T,L}}(l_0,l)\ L^4\
{\delta(P^2) \delta ((R+L)^2) \over R^2 (P+L)^2}.
\label{eq:hardgam}
\ee
Using the
$\delta$-function contraints the term $1 / R^2(P+L)^2$ naively  appears
of ${\cal O}(1/g^2 T^4)$ for $P,\ R$ hard and $L$ soft. However the
denominator $R^2(P+L)^2$ is responsible for collinear divergences which
drastically modify this naive estimate. Using the constraints one 
easily rewrites
\begin{eqnarray}
{-4 \over R^2 (P+L)^2} = {1 \over P\cdot Q} {1 \over P\cdot Q + Q\cdot
L} =
              {1 \over Q\cdot L} \left({1 \over P\cdot Q}
                            - {1 \over P\cdot Q + Q\cdot L} \right) 
                      \approx  {2 \over Q\cdot L} {1 \over P\cdot Q}.
\label{diverg}
\end{eqnarray}
The first equality shows the presence of two very close collinear
singularities (when $P\cdot Q$=0) since the two poles differ only by the soft
$Q\cdot L$ term. The last equality holds true to leading order only
after the integration over the whole phase space is performed.
Introducing the angular variable $u=1-\cos\theta$ between the light-like
momenta $P$ and $Q$ the above expression becomes, near $u=0$,
\begin{eqnarray}
{1 \over R^2 (P+L)^2} \sim  {p \over q L^2}\  {1 \over p q u}
\label{collin}
\end{eqnarray}
This form shows the presence of a logarithmic collinear divergence and
the order of the residue at the pole in $u$ is $1 / g^4 T^4$ instead of
the naively expected $1 / g^2 T^4$. The near overlap of the collinear
singularites causes the enhancement of the predicted rate by a factor
$1/g^2$. The cross section is regularised by keeping the soft kinematic
terms in the hard propagator of the diagrams as well as the hard fermion
thermal mass \cite{FlechR1} which enters the calculation at the same
order. This discussion can be generalised to the case of the production
of quasi-real photons at soft momenta and the relevant parameter is
found to be $Q^2/q_o^2$ \cite{AurenGKP2,Gelis2}: when it is small
($<g^2$) the rate of production is enhanced by a factor $1/g^2$ compared
to the expected order of magnitude in the HTL resummation program while
when $Q^2/q_o^2 \sim 1$ the bremsstrahlung contribution is of the same
order as that estimated in \cite{BraatPY1}. 

The implications of light-cone singularities for the HTL resummation
program present an interesting challenge and they are still to be 
investigated \cite{Rebha2}.

\section{Acknowledgements}
I thank F.~Gelis for many discussions and for much help in the preparation 
of these notes. I also thank the organisers of WHEPP 4 for a very
enjoyable meeting.

%\bibliography{biblio}

\par\section*{Figure Captions}
\begin{itemize}
\item[Fig.~1] The real-time contour.
\item[Fig.~2] Diagram in the resummed theory contributing
to the production of soft photon in a quark-gluon plasma. All momenta 
are soft.
\item[Fig.~3] The various processes involved in the
evaluation of the diagram of the previous figure.
\item[Fig.~4] (a) Processes contributing to hard photon
production in a quark gluon plasma; \\
(b) The corresponding diagram in the Braaten-Pisarski effective theory.
\item[Fig.~5] Diagrams contributing to the gluon damping
rate.
\item[Fig.~6] The dominant diagram contributing to the 
damping rate of a fast fermion; the gluon in the loop is transverse.
\item[Fig.~7] One of the diagram leading to a divergent
contribution in the production of a soft real photon.
\item[Fig.~8] The dominant bremsstrahlung diagrams for 
the production of a soft real photon.

\end{itemize}

%\clearpage

%\vspace*{2.0cm}

\begin{figure}[h]
\label{fig:1}
\vspace*{0.8cm}
\begin{center}
\leavevmode
%\leftskip -1cm
%\vspace*{-2.0cm}
\epsfbox{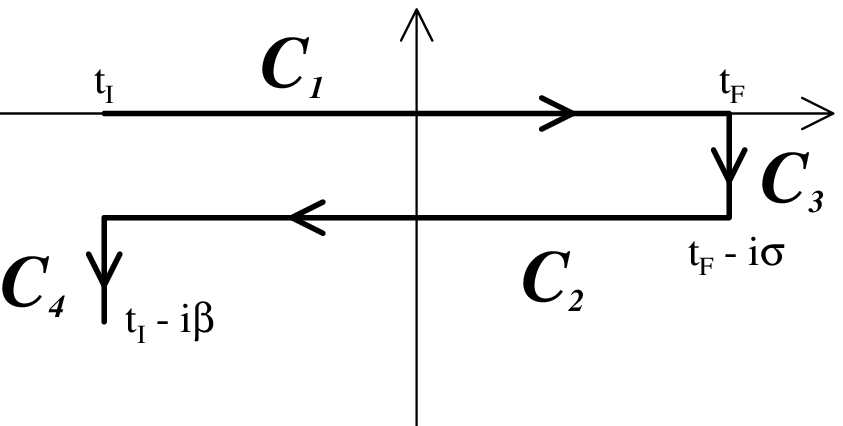}
%\leftskip 0cm
\end{center}
%\caption{Contributions to ...}
\centerline{Fig. 1}
\end{figure}
\vskip -1.0cm
%\centerline{Fig. 1}
%\vskip -0.5cm

\begin{figure}[h]
\begin{center}
\leavevmode
%\leftskip -1cm
\epsfxsize 2.5in
\epsfbox{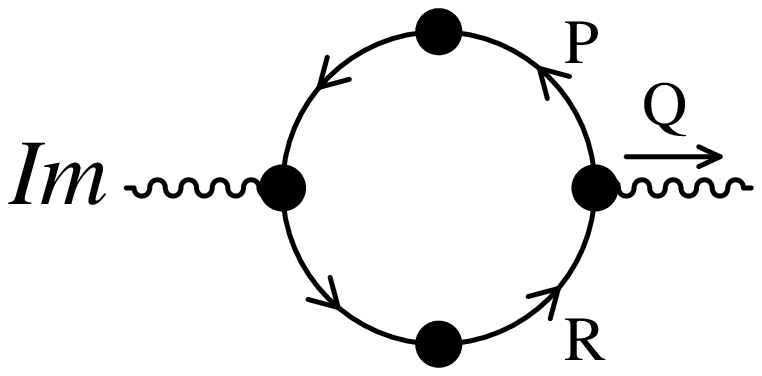}
%\leftskip 0cm
\end{center}
%\caption{Contributions to ...}
\centerline{Fig. 2}
%\label{fig:soft}
\end{figure}
\vskip -0.5cm

\begin{figure}[h]
\label{fig:soft2}
\begin{center}
\leavevmode
%\leftskip -1cm
\epsfxsize 5.5in
\epsfbox{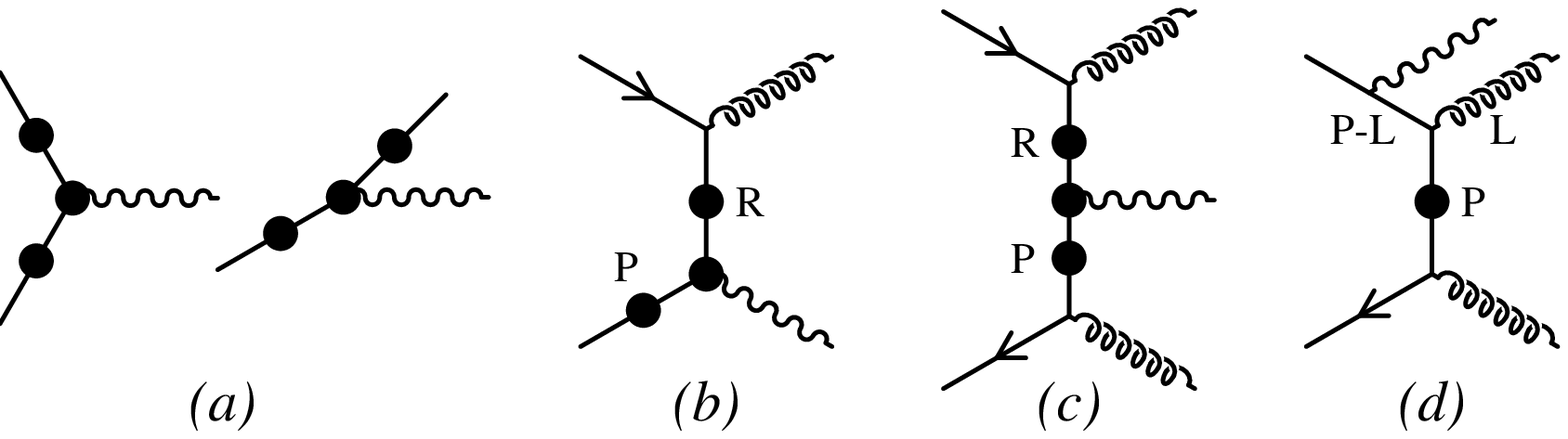}
%\leftskip 0cm
\end{center}
%\caption{Contributions to ...}
\centerline{Fig. 3}
%\vspace*{-10.0cm}
\end{figure}
\vskip -0.5cm

\begin{figure}[hbt]
\begin{center}
\leavevmode
%\leftskip -1cm
\epsfxsize 5.5in
\epsfbox{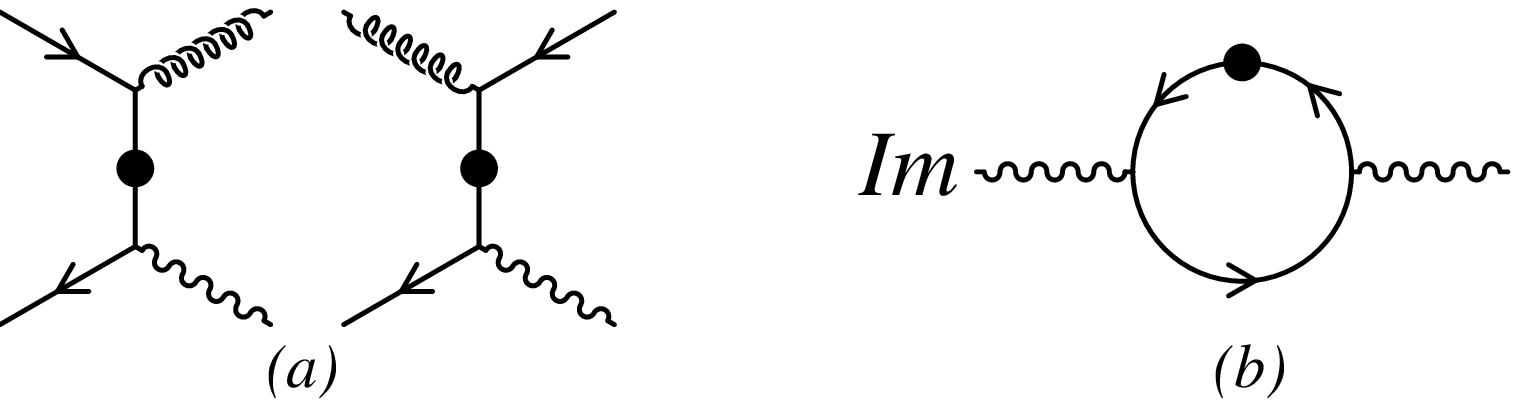}
%\leftskip 0cm
\end{center}
%\caption{Contributions to ...}
\vskip -0.3cm
\centerline{Fig. 4}
%\label{fig:hard}
\end{figure}
%\vskip -0.5cm
%\centerline{Fig. 4}

%\vskip -0.5cm
%\centerline{Fig. 4}
%\begin{figure}[hbt]
%\begin{picture}(198,198)(0,0)
%\put(-10,0){\input{eteta}}
%\vspace*{-2.0cm}
%\vspace*{-2.0cm}
%\put(0,0){\epsfig{file=indfig4.ps,height=3.0cm,width=7.0cm}}
%\end{picture}
%\vspace*{-.5cm}
%\label{fig:hard}
%\end{figure}

%\vskip -0.5cm

\begin{figure}[h]
\begin{center}
\leavevmode
%\leftskip -1cm
%\epsfxsize 5.5in
\epsfbox{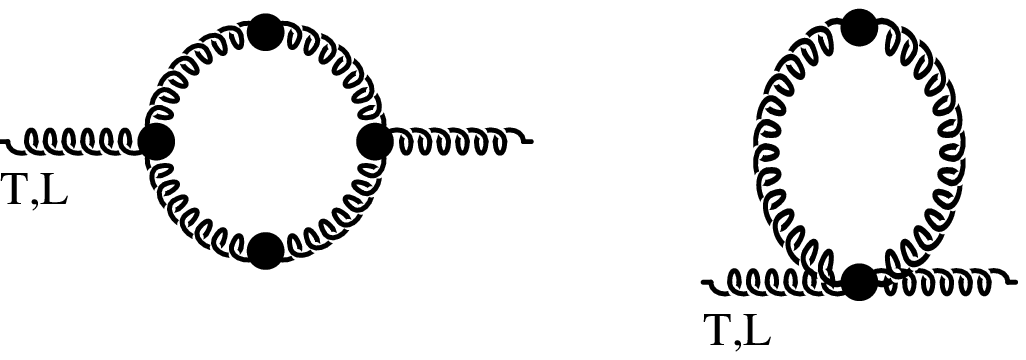}
%\leftskip 0cm
\end{center}
%\caption{Contributions to ...}
\vspace*{-.3cm}
\centerline{Fig. 5}
%\label{fig:damp}
\end{figure}
%\centerline{Fig. 5}

\vskip -0.5cm

\begin{figure}[h]
\begin{center}
\leavevmode
%\leftskip -1cm
%\epsfxsize 5.5in
\epsfbox{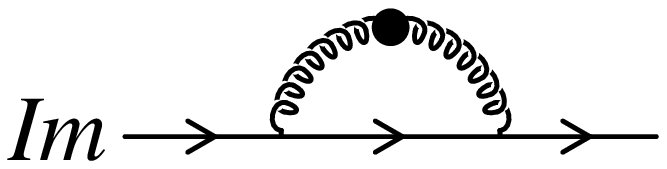}
%\leftskip 0cm
\end{center}
%\caption{Contributions to ...}
\centerline{Fig. 6}
\label{fig:dampf}
\end{figure}
%\centerline{Fig. 6}

\vskip -0.5cm

\begin{figure}[h]
\begin{center}
\leavevmode
%\leftskip -1cm
%\epsfxsize 5.5in
\epsfbox{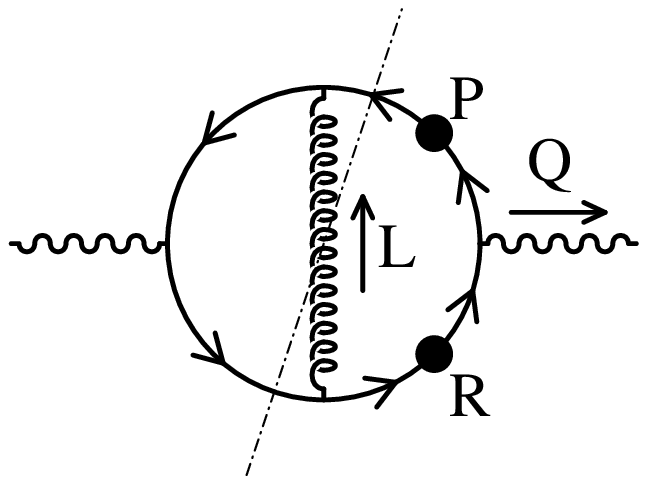}
%\leftskip 0cm
\end{center}
%\caption{Contributions to ...}
\centerline{Fig. 7}
\label{fig:softgam}
\end{figure}
%\centerline{Fig. 7}

\begin{figure}[h]
\begin{center}
\leavevmode
%\leftskip -1cm
\epsfxsize 5.5in
\epsfbox{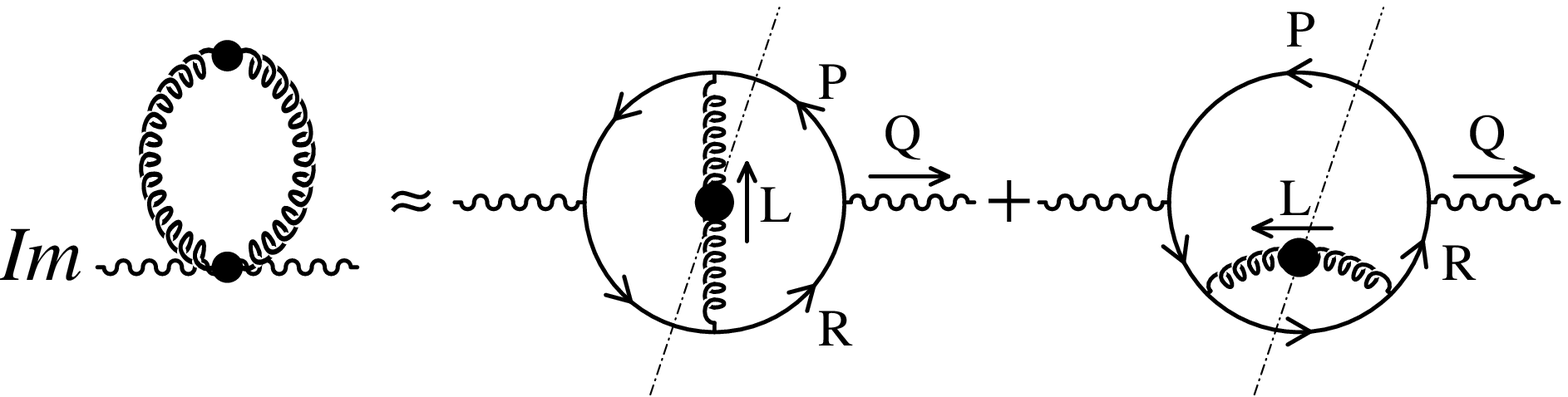}
%\leftskip 0cm
\end{center}
%\caption{Contributions to ...}
\centerline{Fig. 8}
\label{fig:bremgam}
\end{figure}
%\centerline{Fig. 8}

\end{document}